\documentstyle[emulateapj,apjfonts,psfig]{article}
\def\ltsima{$\; \buildrel < \over \sim \;$}
\def\simlt{\lower.5ex\hbox{\ltsima}}
\def\gtsima{$\; \buildrel > \over \sim \;$}
\def\simgt{\lower.5ex\hbox{\gtsima}}
\def \etal {{\it et al.\/}}

\def \cf {{\it cf.\/}}

\def\sec{\hbox{"\hskip-3pt .}}

\def\s{\ifmmode \widetilde \else \~\fi}
\def\={\overline}

\def\spose#1{\hbox to 0pt{#1\hss}}

\def\etal{{\it et al.\ }}
\def\cf{{\it cf.\ }}

\def\lta{\mathrel{\spose{\lower 3pt\hbox{$\mathchar"218$}}
     \raise 2.0pt\hbox{$\mathchar"13C$}}}
\def\gta{\mathrel{\spose{\lower 3pt\hbox{$\mathchar"218$}}
     \raise 2.0pt\hbox{$\mathchar"13E$}}}
\def\Dt{\spose{\raise 1.5ex\hbox{\hskip3pt$\mathchar"201$}}}    
\def\dt{\spose{\raise 1.0ex\hbox{\hskip2pt$\mathchar"201$}}}    

\def\=={\equiv}

\def\dotsfill{\leaders\hbox to 1em{\hss.\hss}\hfill}

\newcommand{\fmmm}[1]{\mbox{$#1$}}
\newcommand{\scnd}{\mbox{\fmmm{''}\hskip-0.3em .}}

\slugcomment{Accepted by The Astrophysical Journal Supplement Series}
 
\lefthead{Ibata et al.}
\righthead{HST photometry of the globular cluster M4}
 
\begin{document}
 
\title{HST Photometry of the Globular Cluster M4}

\author{ 
Rodrigo A. Ibata\altaffilmark{1,2}, 
Harvey B. Richer\altaffilmark{1},
Gregory G.  Fahlman\altaffilmark{1}, 
Michael Bolte\altaffilmark{3}, 
Howard E. Bond\altaffilmark{4}, 
James E. Hesser\altaffilmark{5}, 
Carlton Pryor\altaffilmark{6}, 
Peter B.  Stetson\altaffilmark{5}}
 
\altaffiltext{1}{Department of Physics \& Astronomy, University of British
Columbia, Vancouver, B.C., V6T 1Z4.  E-mail surname@astro.ubc.ca}
 
\altaffiltext{2}
{Present address: European Southern Observatory \nl
Karl Schwarzschild Stra\ss e 2, D-85748 Garching bei M\"unchen, Germany \nl
Electronic mail: ribata@eso.org}
 
\altaffiltext{3}{University of California, Lick Observatory, Santa Cruz, CA
9506 4.  E-mail bolte@ucolick.org}
 
\altaffiltext{4}{Space Telescope Science Institute, 3700 San Martin Drive,
Baltimore, MD 21218. E-mail bond@stsci.edu}
 
\altaffiltext{5}{National   Research   Council,   Herzberg   Institute    of
Astrophysics, Dominion Astrophysical Observatory, 5071 W. Saanich Road, RR5,
Victoria, B.C., Canada V8X 4M6. E-mail firstname.lastname@hia.nrc.ca}
 
\altaffiltext{6}{Rutgers, The State  University of New Jersey, Department of
 Physics  and  Astronomy,  PO Box  849,  Piscataway,  NJ 08855--0849. E-mail
 pryor@physics.rutgers.edu}
 

 
\begin{abstract}
This paper presents a detailed description of the acquisition and processing
of a large body of imaging data for three  fields in the globular cluster M4
taken with the  Wide Field and  Planetary Camera  2 aboard  the Hubble Space
Telescope. Analysis with the ALLFRAME package yielded the deepest photometry
yet obtained   for  this cluster.   The resulting   data-set  for 4708 stars
(positions  and  calibrated photometry  in  V,  I, and,   in two  fields, U)
spanning approximately six cluster core radii is available on the AAS CD-ROM
(or email a request  to RAI).  The scientific analysis  is deferred to three
companion  papers, which investigate the  significant white dwarf population
discovered and the main sequence population.
\end{abstract}

 
\keywords{globular clusters, photometry, white dwarfs, red dwarfs}
 
 
%

\section{Introduction}

The refurbished Hubble Space Telescope (HST) made it possible, for the first
time, to study two  stellar populations that  had remained almost completely
unexplored from the  ground  even  in the   nearest  globular clusters:  the
low-mass red dwarfs at the faint end of the main sequence (MS) and the white
dwarfs (WDs).  To understand in detail the physics of  these objects, and in
particular to constrain  theories of the cooling of  WDs and theories of the
structure of faint MS  stars, we undertook a  study of the  nearest globular
cluster,  M4.  The scientific results of  this study are  presented in three
companion papers: the constraints placed on the  WD population are discussed
in \markcite{Richer95}    Richer  {\it  et   al.}   (1995)   (paper~1)   and
\markcite{Richer97}  Richer  {\it   et al.}    (1997)  (paper~2),  while the
investigation of the MS population is presented in \markcite{Fahl98} Fahlman
{\it et al.}  (1998) (paper~3).  The present paper details the observational
data reductions performed for this study, and forms the basis from which the
analysis of papers~1--3 was extracted.

\section{Observations}

The  refurbished HST  was  used in cycle~4  (8--9  February 1995; 15--16, 29
March 1995;  4--5,  7 April  1995) to   image three fields  in  the globular
cluster  M4 (HST proposal   identification  number 5461).  The  fields  were
chosen such that in the first field the ``Planetary Camera'' (PC) section of
the ``Wide Field and  Planetary Camera'' (WFPC2) mosaic  was centered at the
J2000 coordinates ($\alpha=16^\circ 23' 40''$, $\delta=-26^\circ 31' 30''$),
approximately  at  the center   of M4;  the   second field  was centered  at
($\alpha=16^\circ 23' 42''$,  $\delta=-26^\circ  30' 48''$)  about one  core
radius ($=50\scnd1$ ---  \markcite{Djor93} Djorgovski 1993) from the cluster
center; and  the  third field was  centered at  ($\alpha=16^\circ 23' 56''$,
$\delta=-26^\circ  32' 27''$) approximately  six core radii from the center.
Below, and in the  accompanying papers, we   refer to these fields  as `f0',
`f1'  and `f6',  respectively.  In  Figure~1, the  positions  of these WFPC2
fields are  shown superimposed on a  ground-based image of M4  obtained from
the Digital Sky Survey.

The cluster   was observed  through   the filters   F336W, F555W  and  F814W
(approximately Johnson U and V  and Kron-Cousins I) in  fields  0 and 1;  in
field 6 only F555W and F814W images were obtained.   The total of 28.4 hours
granted for this project was divided  into exposures as listed in Tables~1a,
1b  and 1c.  To  improve the sampling of  the stellar point spread functions
(PSFs),  all repeated  exposures were    ``dithered'',  that is,  they  were
observed at positions slightly offset by non-integer pixel shifts from other
exposures in  the  same field;  these  offsets were  typically  less than 10
pixels   in the   WF   frames. No two frames   were   exposed with identical
pointings.

\section{Data Processing}

Table 2 summarizes the WFPC2  detector characteristics for the observations.
The  images  were  ``pipeline processed''  by  the Space   Telescope Science
Institute:     the    direct current bias    and     dark  current   of  the
charge-coupled-device (CCD)  detectors were   subtracted,  the images   were
flat-fielded, and ``hot-pixels'' (defective regions) on the CCD were flagged
as   unusable.   Next, the image   sections  heavily affected  by vignetting
(typically a 50 pixel  wide band along  the two edges of  the chips that are
closest to the mosaic   center) were masked out.   Finally  all of the  data
frames were multiplied by  a frame (obtained from  the STScI) whose  entries
are the relative pixel  areas over WFPC2; this  corrects for the non-uniform
area of the sky subtended by the pixels over the image plane of the camera.

The  data  were processed  using  the  following photometry  data  reduction
packages:  DAOPHOT   (Stetson  1987  \markcite{Stetson87}),  which  provided
positions  and initial photometry for the  cluster stars; DAOMASTER (Stetson
1992\markcite{Stetson92}) and MONTAGE2  (Stetson  1994\markcite{Stetson94}),
which  were   used  to determine    the  geometrical transformations between
overlapping  frames    and to  combine   them   to  produce  a   clean, high
signal-to-noise  image;  and   ALLFRAME (Stetson  1994\markcite{Stetson94}),
which was  used   to derive the    final point-spread-function (PSF)  fitted
photometry.  In the sections below we detail these reductions.

\subsection{Finding Stars}

The principal   objective of this  experiment was   to   obtain the  deepest
possible accurate photometry of the  cluster main and white dwarf sequences.
The fields studied, especially those close to the  cluster center, display a
very non-uniform sky background.  This  is due in part  to the crowding, but
most importantly  to   the  presence of  very   bright  stars  which produce
highly-saturated images whose extended wings cover a significant fraction of
the frame.     Artificial  star  tests  showed   that this  non-uniform  sky
background was the main  cause of  the  difficulty in detecting faint  local
intensity peaks above the sky, thereby limiting the depth of the photometry.

Therefore, to improve the probability of finding the faint stars, all of the
F555W and F814W exposures in a field were combined to form a median-filtered
frame.  The F336W frames were combined separately,  but in a similar manner.
These combined  frames were obtained as follows.   First, the  routines FIND
and PHOT  within the package DAOPHOT  identified  and measured the positions
and magnitudes of   stars $10 \sigma$  above the  mean  local sky background
noise.  The program DAOMASTER then  yielded a six-coefficient transformation
relating the position of stars  on each frame  to those on a chosen ``master
frame''; it also calculates the average magnitude  offset between each frame
and the  master frame.  Finally,  the  program MONTAGE2 median filtered  and
combined  the frames using   these geometrical transformations and magnitude
offsets.

Figure 2a  is a reproduction of the  combined F555W and  F814W WF3 frames in
the central field (f0), while  Figure 2b shows  the combined F555W and F814W
WF2 frames in  the outermost  field  (f6).  The difficulty  of finding faint
stars in the image shown in Figure 2a is evident: there are long diffraction
spikes criss-crossing the image that emanate from saturated images of stars,
the  sky background  level  increases noticeably towards  the center  of the
cluster, much of the   frame is covered   with charge overflow  columns from
highly-saturated stars,  and  the  ``wings'' of  the   PSFs of bright  stars
contaminate a significant fraction of the frame.

The  PSF varies considerably over  the WFPC2 field;   near the center of the
mosaic  the image   profiles are round,  but, towards   some corners of  the
camera, the image profiles become  significantly elongated.  This means that
one  cannot impose stringent  roundness criteria to differentiate stars from
galaxies and noise  peaks  without losing a  significant amount  of valuable
data.  With  lax roundness criteria,   the standard FIND routine in  DAOPHOT
gave very  many false detections  along diffraction spikes,  since the image
profile of  such  brightness enhancements  often  appear  stellar.  However,
diffraction  spikes always appeared  on diagonal lines  on the  CCDs, so the
roundness parameter  in  FIND was  altered to  check for four-fold symmetry,
rather than simply checking the ratio of  the widths of  the image along the
row   and column  directions.   (The   latter  is  useful   for  eliminating
charge-overflow columns and rows, but did not eliminate linear features at a
45-degree angle). No attempt was made to reject possible spurious detections
that might have occured at  the intersections of  Airy disks; the broad-band
filters  used produce  low contrast   rings,  to which the FIND   algorithm,
judging from artificial star tests, was not sensitive.

After  much experimentation with    the roundness, sharpness  and  threshold
parameters in  the   FIND routine,  we concluded that    we  were unable  to
discriminate sufficiently well against false stellar detections whilst still
retaining most real stars.  Sufficiently far from their sources, many of the
diffraction spikes  fade  into the  sky background;  however, where two such
spikes cross over, the combined intensity  can be raised above the detection
threshold and  a false detection  ensues.  Though the diffraction spikes are
below the sky noise in the immediate neighborhood of the false detection, it
can nevertheless be blatantly obvious by visual inspection of the image that
the  detection   has occurred at   the point  of intersection of   a pair of
diffraction spikes.  We decided therefore to perform a second classification
by examining visually each of the candidate stars in the FIND output list (a
script  was   used to  invoke  the   IRAF routine   IMEXAMINE).  Stars  were
classified as either ``clearly a star'',  ``undecided'' or ``certainly not a
star''  according to their positions with  respect to diffraction spikes and
according  to the shapes  of  their profiles (the  visual inspection  of the
profiles was especially useful where a  detection occurred close to a bright
star: in the strongly sloping background diffraction spikes can be difficult
to trace in a grayscale image, but are easily seen in a  3-D flux map).  All
candidate positions  deemed  visually to be  ``certainly  not a  star'' were
removed  from the lists.   Note that this   selection was very conservative:
very few candidate stars were actually discarded (typically less than 10\%),
and  this  classification   was   performed {\it  before}  any   photometric
measurements were made.

The FIND parameters  were set as  follows (see  \markcite{Stetson87} Stetson
1987 and the DAOPHOT  manual for a  detailed explanation of the parameters):
the readout noise and gain  values of the  detector were taken from Table~2;
the low  data limit was  set  to 10 standard deviations   below the mean sky
value (a larger than usual value being  necessary due to the fairly variable
sky background);  the high data  limit was set to  4090~DN; the low and high
sharpness values were set to 0.2 and 1.0, respectively; and the low and high
roundness values were set to -0.5 and 0.5,  respectively.  The improved FIND
routine described above was set to find $5 \sigma$ peaks above the local sky
background in the combined F555W and F814W frames  and in the combined F336W
frames.

Finally, the star  lists resulting from  the combined F555W and F814W frames
and from the combined F336W frames were merged  to produce a ``master list''
for each field (the lists were summed and multiple entries --- defined to be
detections on  the F336W frame  which were less   than 0.75 pixels  from the
position  of a detection  in the  combined  F555W and F814W frames  --- were
removed).   Some artefacts of the  frames will have   given rise to spurious
detections that still  remain in this list.   Without further data, there is
no sure way or removing such objects,  especially towards fainter end of the
data.  We hope  to remove most of  the remaining false detections by fitting
the images to a  PSF. It is also  very difficult  to provide a  quantitative
estimate of the number of such  spurious detections that  will remain in the
final photometry list.  We can only remark that it  seems unlikely that they
would have colors  that would mimic the main-  and white dwarf sequences in
the cluster.

Note that the median combined frames  were only used  to improve the finding
statistics.  The final photometry was not derived from these frames, but was
carried out on  the  individual un-shifted frames   as we detail in  the next
section.

\subsection{ALLFRAME Reductions}

The     photometry  data    reduction  package    ALLFRAME    was  developed
(\markcite{Stetson94} Stetson  1994) to  make optimal   use of the  sampling
information  contained  in several (slightly  offset)  exposures of  a given
field.   This  routine  fits a   PSF  model  to   all  stars  in all  frames
simultaneously,  and therefore requires a  knowledge of the instrumental PSF
(and  its variation   over  the   frames)   together with the    geometrical
transformations between the frames.

Since the WFPC2 instrumental  PSF was believed  to be relatively  stable, we
chose to make  use of a previously constructed  PSF model (Stetson,  private
comm.)   derived  from  WFPC2 observations  in the   F336W,  F555W and F814W
filters  of the globular cluster  Omega Centauri.  Bright, unsaturated stars
were selected to  sample the PSF, modelled as  a ``Moffat'' function  with a
lookup table,     which    allows  for   deviations     from   the  analytic
profile.  Diffraction spikes were thereby included  in the PSF model, though
no special effort  was  made to improve  the   signal to noise  of  such low
contrast features. As mentioned above, the WFPC2 PSFs vary considerably with
position over  the mosaic and so  were modeled as varying quadratically with
position over each of the  WFPC2 chips. The radius of  the PSF function  was
chosen to be 14 pixels.

ALLFRAME fits the  stellar profiles over  a ``fitting radius''.   After many
experiments it was decided to use a value of 2 pixels for this parameter for
both the PC and WF  data.  Though this compromises  the photometry of bright
saturated  stars, in  trials  it produced the best   photometry of the faint
stars (judging by  the   magnitude error  estimated  by   ALLFRAME).   Other
ALLFRAME parameters  were set as follows:  the minimum and maximum number of
PSF-fitting iterations were 5 and 200, respectively;  the positions of stars
between frames was  allowed to be   refined with a 20 coefficient  geometric
transformation; the inner and outer radius  of the sky annulus  was set to 3
and  15 pixels, respectively; and the  ``clipping exponent'' was  set to the
value 6.   Two further parameters  are used to help  estimate the quality of
the PSF fit:  the ``profile error'' which  was set to  6.0\% (this parameter
defines the  contribution  to the   error due  to  PSF-mismatching), and the
``percentage  error'',   which was   set   to  0.5\% (an  estimate   of  the
contribution from flat-fielding errors).

\subsection{Calibrations}

The   ALLFRAME  routine  derives,  among    other parameters, a   PSF-fitted
magnitude, a magnitude uncertainty and a $\chi$ measurement  for each of the
frames for every star on the ``master list''.

\markcite{Kelson96} Kelson \etal   (1996) have   shown that  the   detection
efficiency of  the   WFPC2 CCDs is    non-linear with exposure  time.   They
recommend applying  a correction of ($+0.05 \pm  0.007$) magnitudes to WFPC2
images which have  in excess of  160e$^-$  in the  mean sky.  We  adopt this
procedure, using  a  linear ramp between  no  correction and +0.05 mags  for
those exposures where the sky counts are less than 160e$^-$.  The sky values
(for the long exposures) in each of the frames are listed in Table~3.

The     instrumental magnitudes  from the  different    frames were combined
separately for    each   filter as  follows.    First,   the magnitudes  and
uncertainties were converted into detected counts per second.  Averaging the
flux measurements  for a given  star and weighting  by the inverse-square of
the uncertainties yielded the mean detected counts for that star.  Then, any
measurement deviating  from this mean by  more than four  times the weighted
mean    error was rejected.    This   process  was  iterated  until  no more
measurements   were rejected  or until    a minimum  number of  measurements
remained. This minimum number of measurements was set to 4  for F336W, 8 for
F555W and 4 for  F814W.  Finally, the  weighted  mean counts  were converted
into instrumental magnitudes.

These averaged  instrumental PSF-fitted magnitudes were calibrated following
the  recipe  of  \markcite{Holtz95b}  Holtzman \etal    (1995b).  First, the
aperture   correction was calculated   for   each  filter  and field;   this
correction  is to account   for the fact  that \markcite{Holtz95b}  Holtzman
\etal (1995b) measure  the   flux from  their standard  stars  with  a $1''$
diameter aperture for both  PC and WF data.  Since  the M4 star  fields were
rather crowded, aperture corrections could not be reliably calculated in the
usual way.  Instead, we  constructed artificial frames using the ``mknoise''
utility in  IRAF  with identical sky  mean  and variance as  the data frames
(these frame  statistics were  measured  with the  SKY routine  in DAOPHOT).
Using the   DAOPHOT  routine  ADDSTAR  with  the  above-mentioned PSF  as  a
template,  49 artificial stars were  added onto the frame over  a range of 5
magnitudes in intensity up  to just below  the  saturation limit.   Aperture
magnitudes for these artificial stars were calculated  with a $1''$ diameter
aperture using the DAOPHOT routine  PHOT, while PSF magnitudes were obtained
using ALLFRAME with identical  parameters as used for   the real data.   The
resulting aperture corrections (in  the sense $M_{PSF} - M_{APERTURE}$)  are
listed  in Table  4; these  values are   added to the  observed instrumental
magnitudes.  A check  of  this  somewhat  unconventional way of    computing
aperture    corrections was made  in    field  f6,  where  we  compared  the
aperture-corrected  PSF magnitudes  with  actual aperture  magnitudes.   All
isolated stars  in f6 (no neighbors within  20 pixels) with F555W  and F814W
PSF magnitudes brighter  than 25  were  chosen for this  test.  The aperture
magnitudes  of each star   were combined in an   identical manner to the PSF
magnitudes.  The F555W PSF magnitudes were found to be, on average, $0.012$,
$0.008$,  $0.005$, and $0.01$  magnitudes   fainter than the F555W  aperture
magnitudes in, repectively, PC1,  WF2, WF3, and  WF4.  An identical test for
the F814W PSF magnitudes show  that these are,  on average, $0.007$, $0.01$,
$0.009$,  and $0.002$ magnitudes fainter than  the F814W aperture magnitudes
in, repectively, PC1,  WF2, WF3, and WF4.   This indicates that the aperture
corrections are reliable at the $\sim 0.01$ magnitude level.

F336W magnitudes must be corrected for the slow degradation in UV throughput
since the last decontamination   date prior to observation.    (Contaminants
slowly accumulate on the CCD windows, reducing the transmission; this effect
is  reversed  approximately once   a  month when  the  CCDs are deliberately
heated).    For     an   operating  temperature   of     ${\rm-88^\circ C}$,
\markcite{Holtz95b} Holtzman  \etal   (1995b) find    that the   the   F336W
throughput  decreases by $0.0007 \pm  0.0005$ magnitudes per  day for the PC
and 0.0026 magnitudes per day for WF2 (we assume this contamination rate for
all of  the  WF  chips).  The  decontamination date  prior  to the  f0 F336W
observations was  March 11th 1995  (24  days elapsed), so the   f0 PC and WF
instrumental F336W magnitudes were corrected  by 0.017 and 0.062 magnitudes,
respectively; while decontamination date prior to  the f1 F336W observations
was January  13th 1995 (26 days elapsed),  so the f1  PC and WF instrumental
F336W magnitudes were corrected by 0.018 and 0.068 magnitudes, respectively.
Note that the  uncertainty in these corrections is  large, contributing to a
systematic $\sim 0.1$ magnitude uncertainty in the F336W photometry.

A correction for the charge-transfer-efficiency (CTE) of the CCD is applied;
charge is lost during the  readout of the CCDs such  that stars at high  row
numbers appear fainter  than  if they are  observed  at low row numbers.  We
correct  for this  effect by  multiplying   observed  counts by  the  factor
$0.02*(y-1)/799.0+1$, where $y$ is the row number.
\footnote{One of us (PBS) has been involved in the calibration of WFPC2 data
for the H$_0$ Key Project. An analysis of those  data in 1996 suggested that
the above procedure for CTE  correction was  appropriate for exposures  with
background  levels similar to the M4  dataset.  Holtzman \etal\ (1995a) find
the same correction for frames with background levels of ${\rm \sim 30 e^-}$
to ${\rm \sim 250  e^-}$.   After the  completion  of the present work,  the
study  of Whitmore \& Heyer  (1997) appeared on  the HST web-site; they find
that charge is  lost along both the row  and column  directions, and provide
formulae for the correction of this artefact.  However, we have opted not to
apply this  new model  for CTE  correction, as  the purpose  of  the present
contribution is to detail the data reduction procedure used in the companion
papers~1,  2  \& 3.     The maximum  difference  between   the  adopted  CTE
correction, and  that suggested by Whitmore \&  Heyer  (1997) is $\sim 0.02$
magnitudes.   A systematic  error of   $\sim  0.01$ magnitudes and an  extra
scatter of $\sim 0.01$ magnitudes  can be expected from this mis-correction,
as the M4 stars are approximately uniformly distributed over each chip.}

The standard stars used by  \markcite{Holtz95b} Holtzman \etal (1995b)  were
observed  with a gain  level setting of 14 (see,  e.g., the WFPC2 handbook).
The observed counts are therefore  divided by the ratio  of the gain setting
used by \markcite{Holtz95b} Holtzman \etal (1995b)  to the gain setting used
in the present  M4 observations (these are  listed  in Table~2 for  the four
WFPC2 chips).

For comparison to photometric standards, the total extinction along the line
of  sight to the program  stars  must be  known.   Following the analysis in
paper~2, we  adopt a reddening value  of ${\rm  E(B-V)=0.37}$.  Because this
value is not negligible, the small dependence of extinction on stellar color
needs to be included  in the correction.   \markcite{Holtz95b}Holtzman \etal
(1995b) tabulate the extinction  in the filters  F336W, F555W and F814W as a
function  of  ${\rm E(B-V)}$ for an  O6  spectrum  and  a K5 spectrum (their
Tables  12A \& 12B).  Interpolating  in these tables for ${\rm E(B-V)=0.37}$
yields:   ${\rm A_{F336W}=1.887}$,   ${\rm   A_{F555W}=1.143}$   and   ${\rm
A_{F814W}=0.659}$  for   an   O6 star  and   ${\rm  A_{F336W}=1.774}$, ${\rm
A_{F555W}=1.101}$   and  ${\rm  A_{F814W}=0.652}$   for a     K5  star.  The
instrumental magnitudes were chosen such that (in all three filters) $m = 25
- 2.5 \log(DN/s)$. In this instrumental system,  we estimated that a O6 star
has ${\rm F336W - F555W = 2.0}$ and ${\rm F555W - F814W =  0.5}$, while a K5
star  has ${\rm F336W  - F555W = 6.5}$  and ${\rm F555W -  F814W = 2.5}$.  A
linear   relation between instrumental  color   and extinction  in the above
bandpasses was thereby determined. The contribution  to the total systematic
uncertainty from the uncertainty in the ${\rm E(B-V)}$ value is quite small;
a 0.01 magnitude error in ${\rm E(B-V)}$ gives rise to  an error of $<0.001$
magnitudes in U, V, and I.

In the case where  the star has measured F336W,  F555W and F814W magnitudes,
the ${\rm F555W - F814W}$ color is used to estimate  the extinction in F555W
and F814W and  the  ${\rm F336W -  F555W}$   color is used to  estimate  the
extinction in F336W.

Finally,     the         synthetic  transformations       from  instrumental
extinction-corrected F555W and F814W  magnitudes to Johnson ${\rm  V_0}$ and
Kron-Cousins  ${\rm I_0}$ (\markcite{Holtz95b}  Holtzman  \etal 1995b, their
Table~10)  and observed transformations   from instrumental F336W to Johnson
${\rm U_0}$  are applied.  The  quoted  uncertainty in  the Holtzman  \etal\
transformations from F555W and  F814W  magnitudes to  V and I  depends quite
sensitively  on color.    For stars  with  ${\rm  V-I  \sim  1}$, the quoted
uncertainty  in the transformations   is relatively small, $\sim  0.002$ and
$\sim 0.004$ magnitudes,  in, respectively, V and  I. However, for red stars
with ${\rm V-I \sim 3}$,  this uncertainty is much  larger, $\sim 0.004$ and
$\sim 0.01$ magnitudes, in, respectively, V and I.  The uncertainties in the
transformation  from   F336W magnitudes  to    Johnson  ${\rm U}$  are  also
color-dependent.   Stars with  ${\rm  U-I  \sim  0}$, are  expected to  have
magnitudes uncertain to $\sim  0.01$.   This uncertainty increases  to $\sim
0.08$ magnitudes at ${\rm U-I \sim 5}$.

The substantial overlap between  fields~0 and 1  allow an internal check  on
the  accuracy of the photometric zero-point.   This is a useful exercise, as
the point-spread function varies substantially over WFPC2, and in particular
it allows a comparison between the PC and WF data, where the pixel scale and
sky are significantly different.  The difference between the 3-sigma clipped
mean magnitudes for all stars with photometric  uncertainties $< 0.1$~mag in
the two fields  is: ${\rm \overline{U_{f0}}  - \overline{U_{f1}} =  0.006}$,
${\rm   \overline{V_{f0}}  -   \overline{V_{f1}}   =   0.012}$   and   ${\rm
\overline{I_{f0}} - \overline{I_{f1}}  = -0.009}$.  Note  that there  may be
some  positional mismatches,  so the quoted  mean  differences are  an upper
limit to the internal systematic errors. Also, any  PSF mismatch cannot give
rise to a systematic error greater than these values.

To summarize, the main sources of systematic error in the V and I photometry
arise  from   the  uncertainty in  the   Kelson  \etal\ correction   for CCD
non-linearity   ($0.007$  mags),  from  the   uncertainty   in the  aperture
corrections ($\sim  0.01$ mags), from the uncertainty  in the CTE correction
($\sim 0.01$ mags), from the  uncertainty in the photometric transformations
($\sim 0.005$   mags), and from any PSF   mismatch ($\simlt 0.01$  mags).  A
reasonable estimate of the systematic uncertainty  in the V and I photometry
is   therefore $\sim 0.04$ magnitudes.    The  systematic uncertainty in the
U-band photometry is  much larger; as discussed  by Holtzman \etal\ (1995a),
the response from chip-to-chip and over a single chip is variable, there are
variable  contamination rates,  and  there  are large  uncertainties in  the
photometric  transformations.  Holtzman \etal\  estimate that UV photometry,
calibrated by following  their recipe, will  be systematically  uncertain to
several tens of percent.

\section{Results}

The complete  calibrated photometry can  be found on  the AAS  CD-ROM in the
form of a 20-column ASCII table.  To clarify the layout of the CD-ROM table,
the  printed version   of  this paper displays  the   first 10 rows  of data
separately in Table~5a (columns 1 to 8)  and in Table~5b  (columns 9 to 20).
The entries in  these two tables are  as  follows.  In  Table~5a, column (1)
lists  the sequential id  number for  the star,  column  (2) lists the field
number (as defined  above), column (3) gives the  chip number (where a value
of 1 denotes the PC chip, and values 2, 3 and 4  denote the WF1, WF2 and WF3
chips, respectively), column  (4) gives the   visual classification id  (\cf
Section~3.1, where the value 0 denotes that the object is ``clearly a star''
and the value  1 denotes ``undecided''), columns  (5) and (6) give the (x,y)
pixel position of the object in  the coordinate system of the first-observed
F814W frame in that field, and columns (7) and (8)  list the right ascension
and  declination (J2000) of the  object (as determined  from the STSDAS task
METRIC).  Table~5b lists the next 18 columns of the CD-ROM table, which give
photometric  quantities,  in three  groups  of six columns.  For  the F336W,
F555W  and F814W data,   respectively, columns (9),  (13)  \& (17)  list the
instrumental  magnitude; columns  (10), (14)   \&  (18) list the  calibrated
magnitude in the Johnson or Kron-Cousins system; columns  (11), (15) \& (19)
give the rms  scatter of the individual measurements  that  were combined to
obtain the  calibrated magnitude; and  columns (12),  (16) \& (20)  list the
mean $\chi$ value of the PSF fit to the object.  Magnitude values of 999.999
or magnitude error and $\chi$ values of 99.999 denote corrupt or unavailable
data.

Figure 3 shows the calibrated  color-magnitude diagram (CMD)  of M4 in V  vs
${\rm V-I}$  (panel a) and  U vs ${\rm U-I}$   (panel b); only  objects with
magnitude  uncertainties less  than 0.5, $\chi$  less than  2, and that were
also classified visually as stars are plotted.   The cluster MS extends from
the bright limit of the data at ${\rm V \sim 18}$, ${\rm V-I \sim 0.9}$ down
to ${\rm V \sim  27}$, ${\rm V-I \sim 3.0}$,  by  which point it  has almost
vanished;  these  stars  are  studied in paper~3.   The  bulge  and  halo MS
populations are visible   as a wide band parallel   to the cluster   MS, but
approximately  3 magnitudes fainter.   The WD cooling sequence  extends from
${\rm V \sim 22}$, ${\rm V-I \sim  0.0}$ to ${\rm V  \sim 28.75}$, ${\rm V-I
\sim 2.0}$ and is the subject of papers~1 and 2.

\subsection{Photometric Accuracy}

An external check of the photometric zero-point of the WFPC2 data is made by
comparing  the HST  data  to ground-based  photometry (Thompson \etal 1990),
calibrated   with Landolt standards,   which was  acquired  at  the the  Las
Campanas   Observatory (LCO) 2.5~m telescope.   In   Figure~4 we compare the
calibrated photometry of  individual stars from field~6  with the  LCO data.
These diagrams show quite satisfactory agreement between  the two data sets.
There is  some  evidence of   a small   systematic difference  between   the
photometric zero-points  of the two systems amounting  to about 0.03 mags in
$V$ and 0.01 mags in $I$.  Since  it is not clear  whether the ground or the
space  data contain this small error,  we  have not  forced  the two sets to
agree, but have adopted the calibration for  the {\it HST} data  as is.  The
zero-point accuracy we find from these WFPC2  data is thus in fair agreement
with  our estimates above  and with that determined  from, for instance, the
HST H$_0$ key project (Kelson \etal 1996\markcite{Kelson96}).

In Figure~5  we show the magnitude  uncertainties for those stars in Table~5
that have $\chi < 2$ and that were classified  visually as stars. The panels
show, from top  to bottom, the U, V  and  I dependence  of of  the magnitude
uncertainties  as a function of  magnitude.  The rise  in the uncertainty at
bright magnitudes  occurs  because the  PSF is being  fit  to the  wings  of
saturated stars.  Towards  the  faint end of   the diagram, the  uncertainty
rises due to Poisson noise in  the stellar image and  in the background sky.
Stars whose magnitude uncertainty is  much greater than the mean uncertainty
at a particular magnitude are usually close to  bright images or diffraction
spikes.  For completeness, in Figure~6 we show the PSF goodness of fit value
$\chi$ derived   by the ALLFRAME  program (see  Stetson  1994  for a precise
definition of this statistic) for all of the stars in Table~5 with magnitude
uncertainty $\sigma <  1.0$~mag and which  were also classified  visually as
stars.  Again,  the three panels show from  top to  bottom the dependence of
$\chi$  with  magnitude for  the  U,  V and I    photometry.  Note that this
statistic  takes on values  $\sim 1$ for most  of the stars in the data-set,
implying a good fit of the PSF to the stellar image.  Values of $\chi \simgt
2$ usually come from stars  which are close to bright  objects on the frame,
or from galaxies.

\section{Incompleteness Corrections}

Since the  star-finding algorithm was  applied  to combined F555W  and F814W
frames, the probability of finding faint stars  will be a sensitive function
of color. To probe  the recovered fraction as a  function of position in the
CMD, artificial stars were added in the two sequences shown in the left hand
panel of Figure~7, from ${\rm V =  20.41}$, ${\rm V-I =  1.36}$ to ${\rm V =
29.43}$, ${\rm V-I = 3.55}$ and from ${\rm V  = 20.41}$, ${\rm V-I = -0.22}$
to ${\rm  V = 29.43}$,  ${\rm V-I  = 2.0}$, approximating  the cluster  main
sequence and  white dwarf sequences,  respectively.  In each artificial star
simulation, 150 MS stars and 150 WD stars were added to the WF frames and 50
MS stars and 50 WD stars were added to the PC frames. These artificial stars
were  added  to the original   data frames,  {\it   not} the median-filtered
frames.   Ten such sets  of  simulated frames were  produced for  each WFPC2
field,  and reduced in  an identical fashion  to  the data frames (including
visual rejection of bad detections). The right  hand panel of Figure~7 shows
the recovered magnitudes of the  artificial stars. The  scatter at a given V
magnitude is greater for the  artificial WDs than for  the MS stars, because
the latter are brighter in I.  The faintest stars  recovered are MS stars of
${\rm V  \sim  29}$.  The recovered  fraction   of stars as   a function  of
magnitude is given for the WD sequence in paper~2 and for the MS in paper~3.

\section{Conclusions}

We have  presented an  account of   the  reductions of  the  deepest stellar
photometry yet derived  for the  globular  cluster M4.  The resulting  data,
tabulated   in the CD-ROM accompanying this   contribution, show  that it is
possible to obtain accurate photometry   with WFPC2 to $V=28$ in  relatively
crowded  fields.  This  photometric survey  is  used in companion papers  to
study the  main- and white   dwarf sequences with unprecedented  detail and
accuracy.

\section*{Apendix}

Due to  an   oversight, we did  not   include a table  of    the white dwarf
luminosity   function in Richer \etal\ (1997),   the  previous paper in this
series. Table~6 is presented here to amend that oversight. The entire table
is   included in the  AAS  CD-ROM;  on paper, we  list    only the first  10
rows. Column  (1) lists  the annulus number;   column (2)  lists the  V-band
magnitude; column (3)  gives  the completeness correction; column  (4) lists
the completeness--corrected cumulative  luminosity function; and  column (5)
gives the uncertainty on the luminosity function.

\acknowledgments
 
This work was  based  on observations with   the NASA/ESA {\it  Hubble Space
Telescope},  obtained at the  Space  Telescope  Science Institute, which  is
operated by  AURA, Inc.,   under  NASA contract  NAS5-26555.  RAI  expresses
gratitude to the Killam Foundation (Canada) and to the Fullam (Dudley) Award
for  support.   The research  of HBR and   GGF is supported  in part through
grants from the Natural Sciences and Engineering Research Council of Canada,
while   that of MB,   HEB and  CP is  provided   by NASA  through  the grant
GO-05461.01-93A  from  the   Space Telescope  Science   Institute,  which is
operated  by the Associated Universities  for Research  in Astronomy, Inc.,
under NASA contract NAS5-26555.

\eject

\def\sec{{\rm\,s}}

\begin{deluxetable}{cccccccc}
\tablenum{1a}
\tablecaption{Exposure times, field f0}
\tablehead{RA        & DEC        & Filter & Exposure time       & Observation
start      & STScI               \nl
             (J2000) &    (J2000) &        &               (sec) & 
      time &       exposure code}
\startdata
16 23 40.05 & $-$26 31 30.29 & F814W & ~600 & 04 Apr 1995 ~9:29AM & U2HE0101T \nl
16 23 40.05 & $-$26 31 30.29 & F336W & 1300 & 04 Apr 1995 ~9:43AM & U2HE0102T \nl
16 23 40.05 & $-$26 31 30.25 & F814W & ~700 & 04 Apr 1995 10:52AM & U2HE0103T \nl
16 23 40.05 & $-$26 31 30.25 & F336W & 1500 & 04 Apr 1995 11:07AM & U2HE0104T \nl
16 23 40.06 & $-$26 31 30.26 & F814W & ~700 & 04 Apr 1995 12:28PM & U2HE0105T \nl
16 23 40.06 & $-$26 31 30.26 & F336W & 1500 & 04 Apr 1995 12:43PM & U2HE0106T \nl
16 23 40.06 & $-$26 31 30.32 & F814W & ~700 & 04 Apr 1995 ~2:05PM & U2HE0107T \nl
16 23 40.06 & $-$26 31 30.32 & F336W & 1500 & 04 Apr 1995 ~2:20PM & U2HE0108T \nl
16 23 40.06 & $-$26 31 30.42 & F814W & ~700 & 04 Apr 1995 ~3:41PM & U2HE0109T \nl
16 23 40.06 & $-$26 31 30.42 & F336W & 1500 & 04 Apr 1995 ~3:56PM & U2HE010AM \nl
16 23 40.06 & $-$26 31 30.46 & F814W & ~700 & 04 Apr 1995 ~5:18PM & U2HE010BM \nl
16 23 40.06 & $-$26 31 30.46 & F336W & 1500 & 04 Apr 1995 ~5:33PM & U2HE010CM \nl
16 23 40.05 & $-$26 31 30.44 & F814W & ~700 & 04 Apr 1995 ~6:55PM & U2HE010DN \nl
16 23 40.05 & $-$26 31 30.44 & F336W & 1500 & 04 Apr 1995 ~7:10PM & U2HE010EN \nl
16 23 40.05 & $-$26 31 30.38 & F814W & ~700 & 04 Apr 1995 ~8:32PM & U2HE010FP \nl
16 23 40.05 & $-$26 31 30.38 & F336W & 1500 & 04 Apr 1995 ~8:47PM & U2HE010GP \nl
16 23 40.05 & $-$26 31 30.29 & F555W & 1000 & 05 Apr 1995 ~9:35AM & U2HE0201T \nl
16 23 40.05 & $-$26 31 30.26 & F555W & 1000 & 05 Apr 1995 ~9:57AM & U2HE0202T \nl
16 23 40.06 & $-$26 31 30.27 & F555W & 1000 & 05 Apr 1995 11:00AM & U2HE0203T \nl
16 23 40.06 & $-$26 31 30.28 & F555W & 1000 & 05 Apr 1995 11:22AM & U2HE0204T \nl
16 23 40.07 & $-$26 31 30.29 & F555W & 1000 & 05 Apr 1995 12:37PM & U2HE0205T \nl
16 23 40.07 & $-$26 31 30.33 & F555W & 1000 & 05 Apr 1995 12:59PM & U2HE0206T \nl
16 23 40.07 & $-$26 31 30.39 & F555W & 1000 & 05 Apr 1995 ~2:14PM & U2HE0207T \nl
16 23 40.07 & $-$26 31 30.46 & F555W & 1000 & 05 Apr 1995 ~2:36PM & U2HE0208T \nl
16 23 40.07 & $-$26 31 30.52 & F555W & 1000 & 05 Apr 1995 ~3:50PM & U2HE0209T \nl
16 23 40.06 & $-$26 31 30.55 & F555W & 1000 & 05 Apr 1995 ~4:12PM & U2HE020AT \nl
16 23 40.06 & $-$26 31 30.54 & F555W & 1000 & 05 Apr 1995 ~5:27PM & U2HE020BT \nl
16 23 40.05 & $-$26 31 30.52 & F555W & 1000 & 05 Apr 1995 ~5:49PM & U2HE020CT \nl
16 23 40.05 & $-$26 31 30.52 & F555W & 1000 & 05 Apr 1995 ~7:03PM & U2HE020DN \nl
16 23 40.05 & $-$26 31 30.48 & F555W & 1000 & 05 Apr 1995 ~7:25PM & U2HE020EN \nl
16 23 40.05 & $-$26 31 30.41 & F555W & 1000 & 05 Apr 1995 ~8:40PM & U2HE020FN \nl
\hline
Total exposure: &            &       &      &                     &           \nl
            &                & F336W &11800 &                     &           \nl
            &                & F555W &15000 &                     &           \nl
            &                & F814W &~5500 &                     &           \nl
\enddata
\end{deluxetable}

\begin{deluxetable}{cccccccc}
\tablenum{1b}
\tablecaption{Exposure times, field f1}
\tablehead{RA        & DEC        & Filter & Exposure time       & Observation
start      & STScI               \nl
             (J2000) &    (J2000) &        &               (sec) & 
      time &       exposure code}
\startdata
16 23 42.84 & $-$26 30 48.93 & F814W & ~600 & 08 Feb 1995 ~4:48PM & U2HE0301T \nl
16 23 42.84 & $-$26 30 48.93 & F336W & 1300 & 08 Feb 1995 ~5:02PM & U2HE0302T \nl
16 23 42.84 & $-$26 30 48.89 & F814W & ~700 & 08 Feb 1995 ~6:15PM & U2HE0303T \nl
16 23 42.84 & $-$26 30 48.89 & F336W & 1500 & 08 Feb 1995 ~6:30PM & U2HE0304T \nl
16 23 42.85 & $-$26 30 48.91 & F814W & ~700 & 08 Feb 1995 ~7:52PM & U2HE0305T \nl
16 23 42.85 & $-$26 30 48.91 & F336W & 1500 & 08 Feb 1995 ~8:07PM & U2HE0306T \nl
16 23 42.85 & $-$26 30 48.97 & F814W & ~700 & 08 Feb 1995 ~9:28PM & U2HE0307T \nl
16 23 42.85 & $-$26 30 48.97 & F336W & 1500 & 08 Feb 1995 ~9:43PM & U2HE0308T \nl
16 23 42.85 & $-$26 30 49.06 & F814W & ~700 & 08 Feb 1995 11:05PM & U2HE0309T \nl
16 23 42.85 & $-$26 30 49.06 & F336W & 1500 & 08 Feb 1995 11:20PM & U2HE030AT \nl
16 23 42.84 & $-$26 30 49.10 & F814W & ~700 & 09 Feb 1995 12:41AM & U2HE030BT \nl
16 23 42.84 & $-$26 30 49.10 & F336W & 1500 & 09 Feb 1995 12:56AM & U2HE030CT \nl
16 23 42.84 & $-$26 30 49.09 & F814W & ~700 & 09 Feb 1995 ~2:20AM & U2HE030DT \nl
16 23 42.84 & $-$26 30 49.09 & F336W & 1500 & 09 Feb 1995 ~2:35AM & U2HE030ET \nl
16 23 42.83 & $-$26 30 49.03 & F814W & ~700 & 09 Feb 1995 ~3:56AM & U2HE030FT \nl
16 23 42.83 & $-$26 30 49.03 & F336W & 1500 & 09 Feb 1995 ~4:11AM & U2HE030GT \nl
16 23 42.83 & $-$26 30 48.99 & F555W & 1000 & 07 Apr 1995 ~8:38AM & U2HE0401T \nl
16 23 42.83 & $-$26 30 48.96 & F555W & 1000 & 07 Apr 1995 ~9:41AM & U2HE0402T \nl
16 23 42.84 & $-$26 30 48.97 & F555W & 1000 & 07 Apr 1995 10:03AM & U2HE0403T \nl
16 23 42.84 & $-$26 30 48.98 & F555W & 1000 & 07 Apr 1995 11:18AM & U2HE0404T \nl
16 23 42.85 & $-$26 30 48.99 & F555W & 1000 & 07 Apr 1995 11:40AM & U2HE0405T \nl
16 23 42.85 & $-$26 30 49.03 & F555W & 1000 & 07 Apr 1995 12:55PM & U2HE0406T \nl
16 23 42.85 & $-$26 30 49.09 & F555W & 1000 & 07 Apr 1995 ~1:17PM & U2HE0407T \nl
16 23 42.85 & $-$26 30 49.16 & F555W & 1000 & 07 Apr 1995 ~2:31PM & U2HE0408T \nl
16 23 42.85 & $-$26 30 49.22 & F555W & 1000 & 07 Apr 1995 ~2:53PM & U2HE0409T \nl
16 23 42.84 & $-$26 30 49.25 & F555W & 1000 & 07 Apr 1995 ~4:08PM & U2HE040AT \nl
16 23 42.84 & $-$26 30 49.24 & F555W & 1000 & 07 Apr 1995 ~4:30PM & U2HE040BT \nl
16 23 42.83 & $-$26 30 49.22 & F555W & 1000 & 07 Apr 1995 ~5:44PM & U2HE040CT \nl
16 23 42.83 & $-$26 30 49.22 & F555W & 1000 & 07 Apr 1995 ~6:06PM & U2HE040DT \nl
16 23 42.83 & $-$26 30 49.18 & F555W & 1000 & 07 Apr 1995 ~7:21PM & U2HE040ET \nl
16 23 42.83 & $-$26 30 49.11 & F555W & 1000 & 07 Apr 1995 ~7:43PM & U2HE040FT \nl
\hline
Total exposure: &            &       &      &                     &           \nl
            &                & F336W &11800 &                     &           \nl
            &                & F555W &15000 &                     &           \nl
            &                & F814W &~5500 &                     &           \nl
\enddata
\end{deluxetable}

\begin{deluxetable}{cccccccc}
\tablenum{1c}
\tablecaption{Exposure times, field f6}
\tablehead{RA        & DEC        & Filter & Exposure time       & Observation
start      & STScI               \nl
             (J2000) &    (J2000) &        &               (sec) & 
      time &       exposure code}
\startdata
16 23 56.89 & $-$26 32 27.89 & F555W & 2100 & 04 Apr 1995 10:22PM & U2HE0501T \nl
16 23 56.89 & $-$26 32 27.86 & F555W & 2100 & 04 Apr 1995 11:49PM & U2HE0502T \nl
16 23 56.90 & $-$26 32 27.87 & F555W & 2100 & 05 Apr 1995 ~1:24AM & U2HE0503T \nl
16 23 56.90 & $-$26 32 27.88 & F555W & 2100 & 05 Apr 1995 ~2:58AM & U2HE0504T \nl
16 23 56.91 & $-$26 32 27.89 & F555W & 2100 & 05 Apr 1995 ~4:34AM & U2HE0505T \nl
16 23 56.91 & $-$26 32 27.93 & F555W & 2100 & 05 Apr 1995 ~6:11AM & U2HE0506T \nl
16 23 56.91 & $-$26 32 27.99 & F555W & 2100 & 05 Apr 1995 ~7:47AM & U2HE0507T \nl
16 23 56.91 & $-$26 32 28.06 & F555W & 2100 & 29 Mar 1995 ~5:26AM & U2HE0601T \nl
16 23 56.91 & $-$26 32 28.12 & F555W & 2100 & 29 Mar 1995 ~6:52AM & U2HE0602T \nl
16 23 56.90 & $-$26 32 28.15 & F555W & 2100 & 29 Mar 1995 ~8:28AM & U2HE0603T \nl
16 23 56.90 & $-$26 32 28.14 & F555W & 2100 & 29 Mar 1995 10:05AM & U2HE0604T \nl
16 23 56.90 & $-$26 32 28.12 & F555W & 2100 & 29 Mar 1995 11:41AM & U2HE0605T \nl
16 23 56.89 & $-$26 32 28.12 & F555W & 2100 & 29 Mar 1995 ~1:18PM & U2HE0606T \nl
16 23 56.89 & $-$26 32 28.08 & F555W & 2100 & 29 Mar 1995 ~2:54PM & U2HE0607T \nl
16 23 56.89 & $-$26 32 28.01 & F555W & 2100 & 29 Mar 1995 ~4:31PM & U2HE0608T \nl
16 23 56.90 & $-$26 32 27.83 & F814W & ~800 & 15 Mar 1995 ~6:11PM & U2HE0701T \nl
16 23 56.90 & $-$26 32 27.79 & F814W & ~800 & 15 Mar 1995 ~6:30PM & U2HE0702T \nl
16 23 56.91 & $-$26 32 27.81 & F814W & ~800 & 15 Mar 1995 ~7:36PM & U2HE0703T \nl
16 23 56.91 & $-$26 32 27.87 & F814W & ~800 & 15 Mar 1995 ~7:55PM & U2HE0704T \nl
16 23 56.91 & $-$26 32 27.96 & F814W & ~800 & 15 Mar 1995 ~9:18PM & U2HE0705T \nl
16 23 56.91 & $-$26 32 28.00 & F814W & ~800 & 15 Mar 1995 ~9:37PM & U2HE0706T \nl
16 23 56.90 & $-$26 32 27.99 & F814W & ~800 & 15 Mar 1995 11:01PM & U2HE0707T \nl
16 23 56.90 & $-$26 32 27.93 & F814W & ~800 & 15 Mar 1995 11:20PM & U2HE0708T \nl
16 23 56.90 & $-$26 32 27.90 & F814W & ~800 & 16 Mar 1995 12:43AM & U2HE0709T \nl
\hline
Total exposure: &            &       &      &                     &           \nl
            &                & F555W &31500 &                     &           \nl
            &                & F814W &~7200 &                     &           \nl
\enddata
\end{deluxetable}

\begin{deluxetable}{ccccc}
\tablenum{2}
\tablecaption{WFPC2 detector characteristics}
\tablehead{Chip & pixel size & field size & gain & read noise\nl
 & ($''$) & ($''$) & ($e^-/{\rm ADU}$) & ADU }
\startdata
PC1 & 0.046 & $37 \times 37$ & 7.12 & 0.74 \nl
WF2 & 0.1~~ & $80 \times 80$ & 7.12 & 0.77 \nl
WF3 & 0.1~~ & $80 \times 80$ & 6.90 & 0.76 \nl
WF4 & 0.1~~ & $80 \times 80$ & 7.10 & 0.73 \nl
\enddata
\end{deluxetable}

\begin{deluxetable}{ccccccccc}
\tablenum{3}
\tablecaption{Mean sky in frames}
\tablehead{
Chip & \multispan3 f0 & \multispan3 f1 & \multispan2 f6 \cr
  & F336W & F555W & F814W & F336W & F555W & F814W & F555W & F814W}
\startdata
PC1  &  ~~6. & ~56. & ~49.  &  ~~4. & ~48. & ~48.  &  ~60. & ~26.  \cr
WF2  &  ~14. & 154. & 118.  &  ~16. & 165. & 175.  &  296. & 119.  \cr
WF3  &  ~16. & 160. & 119.  &  ~15. & 148. & 154.  &  286. & 116.  \cr
WF4  &  ~16. & 174. & 143.  &  ~19. & 181. & 190.  &  290. & 119.  \cr
\enddata
\end{deluxetable}

\begin{deluxetable}{ccccccccc}
\tablenum{4}
\tablecaption{Aperture corrections}
\tablehead{
Chip & \multispan3 f0 & \multispan3 f1 & \multispan2 f6 \cr
  & F336W & F555W & F814W & F336W & F555W & F814W & F555W & F814W}
\startdata
PC1  &  0.030 & 0.024 & 0.023  &  0.029 & 0.027 & 0.021  &  0.029 & 0.018  \cr
WF2  &  0.057 & 0.047 & 0.047  &  0.060 & 0.040 & 0.046  &  0.037 & 0.050  \cr
WF3  &  0.069 & 0.044 & 0.055  &  0.060 & 0.039 & 0.046  &  0.042 & 0.051  \cr
WF4  &  0.056 & 0.053 & 0.051  &  0.053 & 0.046 & 0.048  &  0.053 & 0.055  \cr
\enddata
\end{deluxetable}

\begin{deluxetable}{cccccccc}
\scriptsize
\tablenum{5a}
\tablecaption{The photometry}
\tablehead{
id & field & chip & class & x & y & RA (2000) & DEC (2000) \cr
(1) & (2) & (3) & (4) & (5) & (6) & (7) & (8)}
\startdata
~1 & 0 & 1 & 0 & 620.99 &  65.79 & 245.90794 & -26.52923 \nl
~2 & 0 & 1 & 0 & 639.38 &  82.59 & 245.90759 & -26.52919 \nl
~3 & 0 & 1 & 0 & 457.95 &  66.56 & 245.90981 & -26.52804 \nl
~4 & 0 & 1 & 0 & ~83.86 &  69.46 & 245.91406 & -26.52529 \nl
~5 & 0 & 1 & 0 & ~85.77 &  75.06 & 245.91400 & -26.52524 \nl
~6 & 0 & 1 & 0 & 339.62 &  76.05 & 245.91109 & -26.52708 \nl
~7 & 0 & 1 & 0 & 761.35 &  81.85 & 245.90620 & -26.53007 \nl
~8 & 0 & 1 & 0 & 357.29 &  83.16 & 245.91083 & -26.52714 \nl
~9 & 0 & 1 & 0 & 523.63 &  83.70 & 245.90892 & -26.52834 \nl
10 & 0 & 1 & 0 & ~99.49 &  95.67 & 245.91368 & -26.52513 \nl
\enddata
\end{deluxetable}

\begin{deluxetable}{cccccccccccc}
\scriptsize
\tablenum{5b}
\tablecaption{The photometry (further columns)}
\tablehead{
${\rm F336W}$ & ${\rm U}$ & ${\rm \sigma_U}$ & ${\rm \chi_U}$ &
${\rm F555W}$ & ${\rm V}$ & ${\rm \sigma_V}$ & ${\rm \chi_V}$ & 
${\rm F814W}$ & ${\rm I}$ & ${\rm \sigma_I}$ & ${\rm \chi_I}$ \cr
 (9) & (10) & (11) & (12) & 
(13) & (14) & (15) & (16) & 
(17) & (18) & (19) & (20)}
\startdata
21.216 & 21.130 & 99.999 & 6.320 &
19.794 & 19.788 & ~0.050 & 1.516 &
18.566 & 18.531 & ~0.090 & 1.256 \nl
25.573 & 27.057 & ~2.062 & 1.244 &
18.006 & 17.994 & ~0.105 & 1.026 &
16.456 & 16.413 & ~0.154 & 1.570 \nl
24.889 & 25.119 & ~1.336 & 1.502 &
19.284 & 19.277 & ~0.027 & 1.164 &
17.995 & 17.959 & ~0.021 & 1.256 \nl
24.579 & 25.002 & ~0.582 & 2.941 &
18.506 & 18.506 & ~0.008 & 1.586 &
17.549 & 17.519 & ~0.000 & 1.080 \nl
24.788 & 25.246 & ~1.362 & 1.893 &
18.641 & 18.637 & ~0.024 & 1.266 &
17.498 & 17.465 & ~0.028 & 0.968 \nl
25.836 & 26.309 & ~0.712 & 1.235 &
19.659 & 19.650 & ~0.035 & 1.234 &
18.269 & 18.232 & ~0.023 & 0.861 \nl
26.035 & 26.364 & ~1.306 & 1.214 &
20.176 & 20.164 & ~0.034 & 1.402 &
18.641 & 18.600 & ~0.017 & 0.952 \nl
25.734 & 26.262 & ~0.717 & 0.942 &
19.449 & 19.440 & ~0.048 & 1.026 &
18.095 & 18.059 & ~0.015 & 0.792 \nl
23.356 & 23.426 & ~0.356 & 1.865 &
18.258 & 18.258 & ~0.098 & 1.029 &
17.303 & 17.274 & ~0.270 & 1.086 \nl
25.894 & 27.140 & ~1.572 & 1.197 &
18.574 & 18.568 & ~0.043 & 1.194 &
17.372 & 17.338 & ~0.017 & 0.914 \nl
\enddata
\end{deluxetable}

\begin{deluxetable}{ccccc}
\tablenum{6}
\tablecaption{Cumulative white dwarf luminosity function}
\tablehead{
Annulus & ${\rm V}$ & completeness & $\Phi$ & $\sigma(\Phi)$
}
\startdata
  1  &  22.64200  &  0.47807   &   2.092   &   0.363 \cr
  1  &  22.89200  &  0.50610   &   4.068   &   0.497 \cr
  1  &  22.89900  &  0.50639   &   6.042   &   0.601 \cr
  1  &  23.11800  &  0.50451   &   8.025   &   0.690 \cr
  1  &  23.30400  &  0.48907   &  10.069   &   0.774 \cr
  1  &  23.37500  &  0.48058   &  12.150   &   0.853 \cr
  1  &  23.67400  &  0.43368   &  14.456   &   0.942 \cr
  1  &  23.67900  &  0.43278   &  16.767   &   1.024 \cr
  1  &  23.77200  &  0.41556   &  19.173   &   1.107 \cr
  1  &  23.88900  &  0.39290   &  21.718   &   1.195 \cr
\enddata
\end{deluxetable}

\eject

\begin{figure}[htb]
\psfig{figure=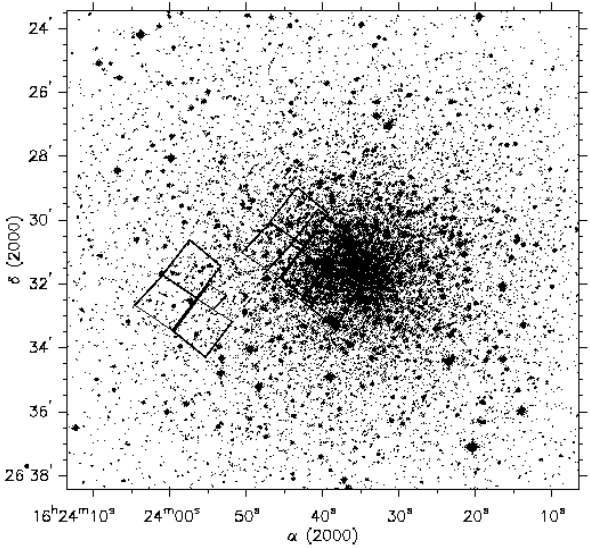,width=8.0cm,angle=0}
\caption[  ]{
The WFPC2 field positions are shown superimposed on  a ground-based image of
M4 extracted from the Digital Sky Survey (second generation).  The positions
of  the vertices  of the  WFPC2 fields were  derived  using the STSDAS  task
``METRIC''. The non-rectangular projection  of the CCDs on the sky is due to
the optical distortion of the WFPC2 camera.}
\end{figure}

\begin{figure}[htb]
\hbox{\psfig{figure=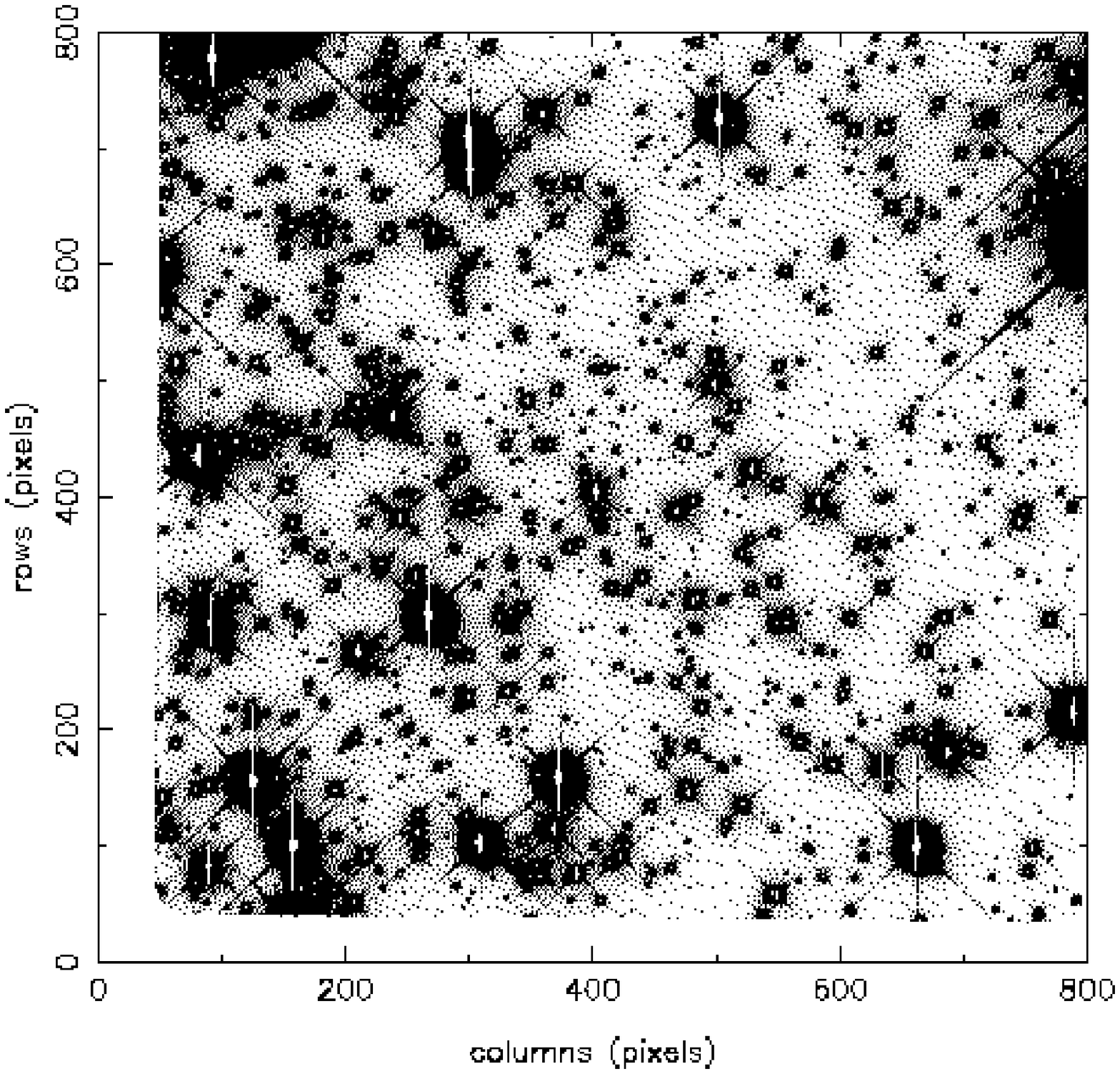,width=8.0cm,angle=0}
\psfig{figure=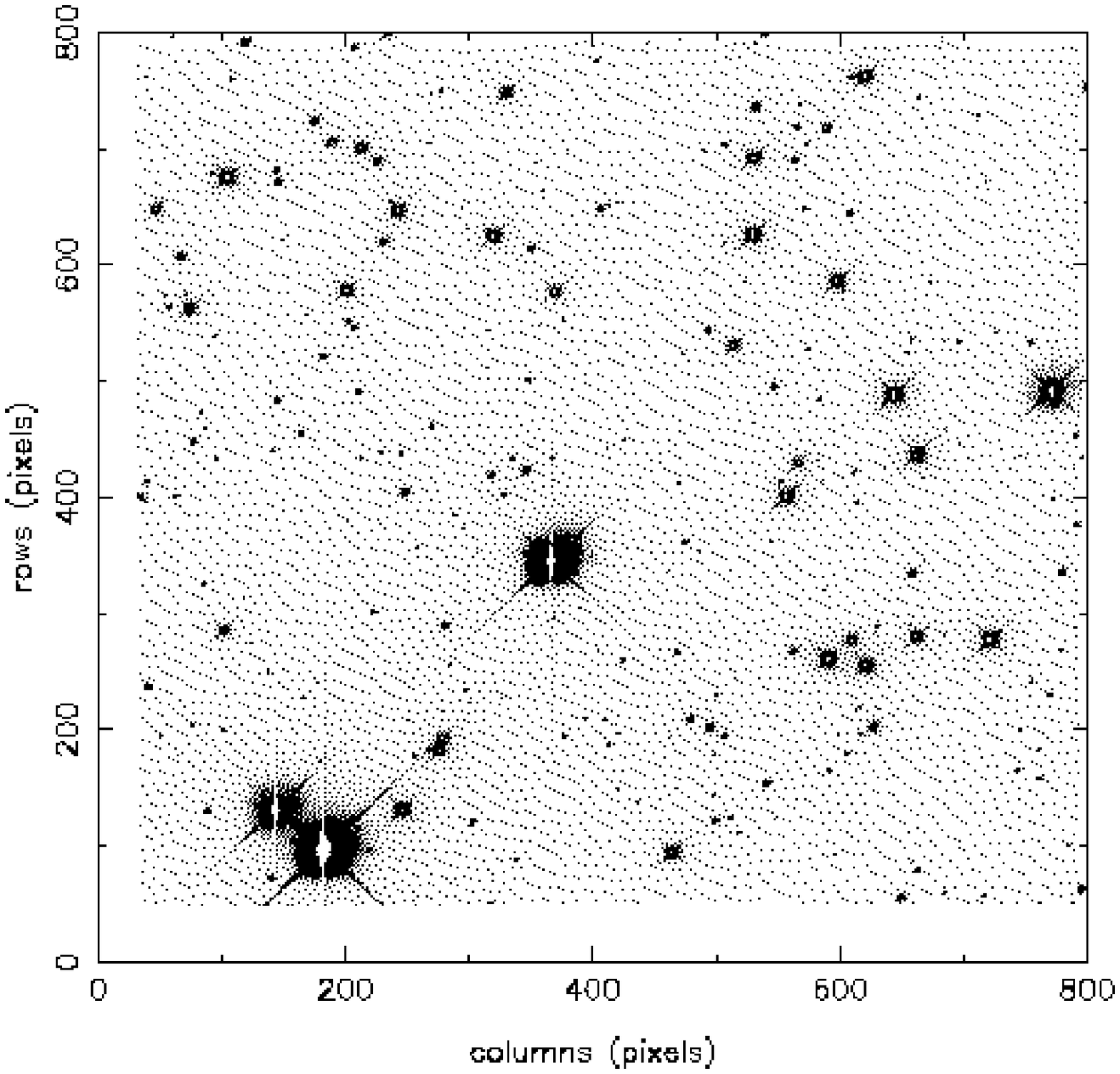,width=8.0cm,angle=0}}
\caption[  ]{
Panel (a)  shows a median-filtered image combined  from all of the available
F555W  and F814W  exposures in  the WF3 chip  in field   f0.  Obtaining good
photometry in such situations is clearly difficult, as the field is crowded,
there are many long  diffraction spikes, the sky  background is not uniform,
and  the diffuse halos  that surround many bright  stars contaminate a large
fraction of the image. For comparison, in panel (b)  we display the combined
F555W and F814W   data from the  WF2 chip  in  field f6,   which yielded the
deepest photometry.  The white  patches on these  representations correspond
to regions where the CCD was saturated.}
\end{figure}

\begin{figure}[htb]
\psfig{figure=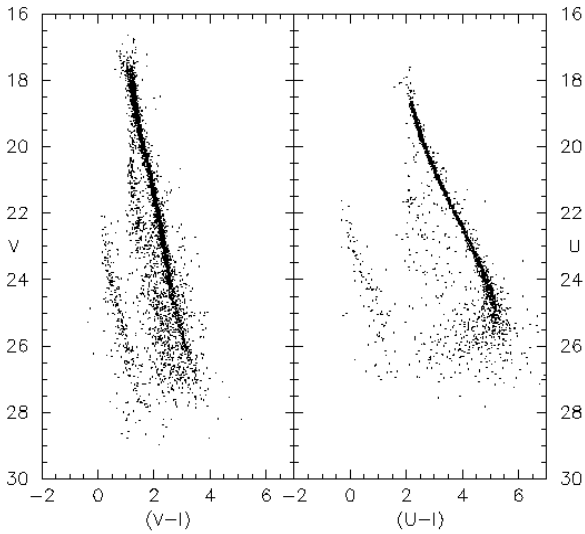,width=8.0cm,angle=270}
\caption[  ]{
The calibrated photometry  is displayed. The  left-hand panel shows the V vs
${\rm V-I}$  CMD containing data from all  of the observed fields, while the
right hand panel gives the U vs ${\rm U-I}$  CMD for fields  f0 and f1.  All
objects  plotted  on these  diagrams have  photometric uncertainties of less
than 0.5 mags, $\chi < 2.0$, and were confirmed visually to be stars.}
\end{figure}

\begin{figure}[htb]
\psfig{figure=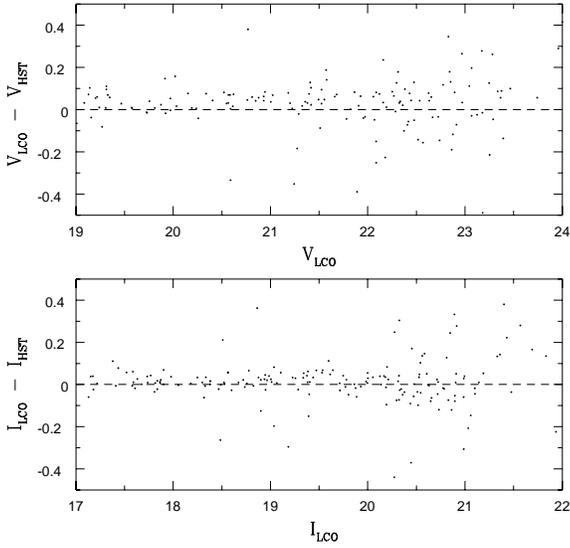,width=8.0cm,angle=0}
\caption[  ]{
A comparison between  ground-based ${\rm V}$ and  ${\rm  I}$ photometry from
the Las Campanas 2.5 meter  telescope and that from  {\it HST}.  All of  the
stars shown are from field~6.}
\end{figure}

\begin{figure}[htb]
\psfig{figure=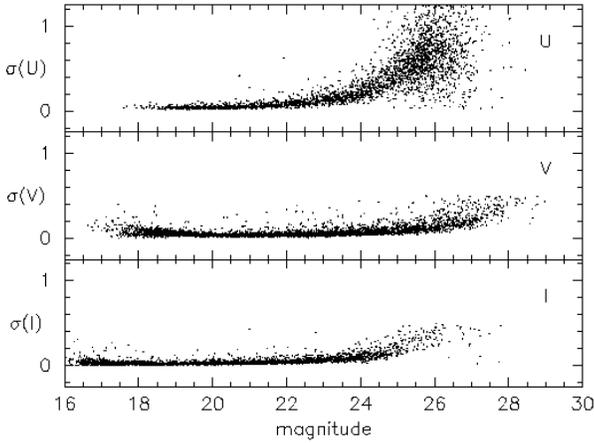,width=8.0cm,angle=270}
\caption[  ]{
The panels show, from top to bottom, the dependence  on magnitude for the U,
V, and I photometry of the magnitude uncertainties  for all stars in Table~5
that have $\chi < 2$ and that were classified visually as stars.}
\end{figure}

\begin{figure}[htb]
\psfig{figure=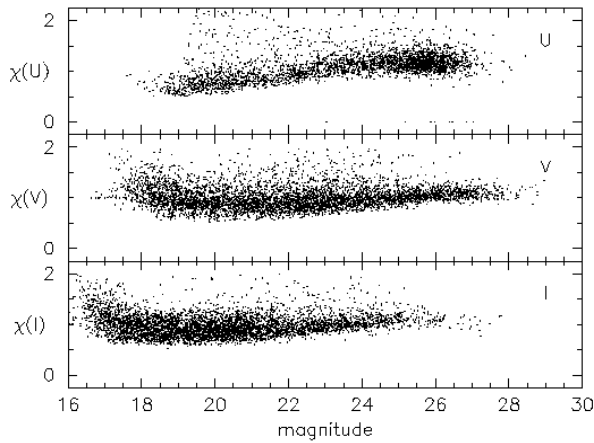,width=8.0cm,angle=270}
\caption[  ]{
The panels show, from top to bottom, the  dependence on magnitude for the U,
V, and I photometry of the $\chi$ goodness of fit statistic for all stars in
Table~5 that  have  magnitude error less than   1 and  that were  classified
visually as stars.}
\end{figure}

\begin{figure}[htb]
\psfig{figure=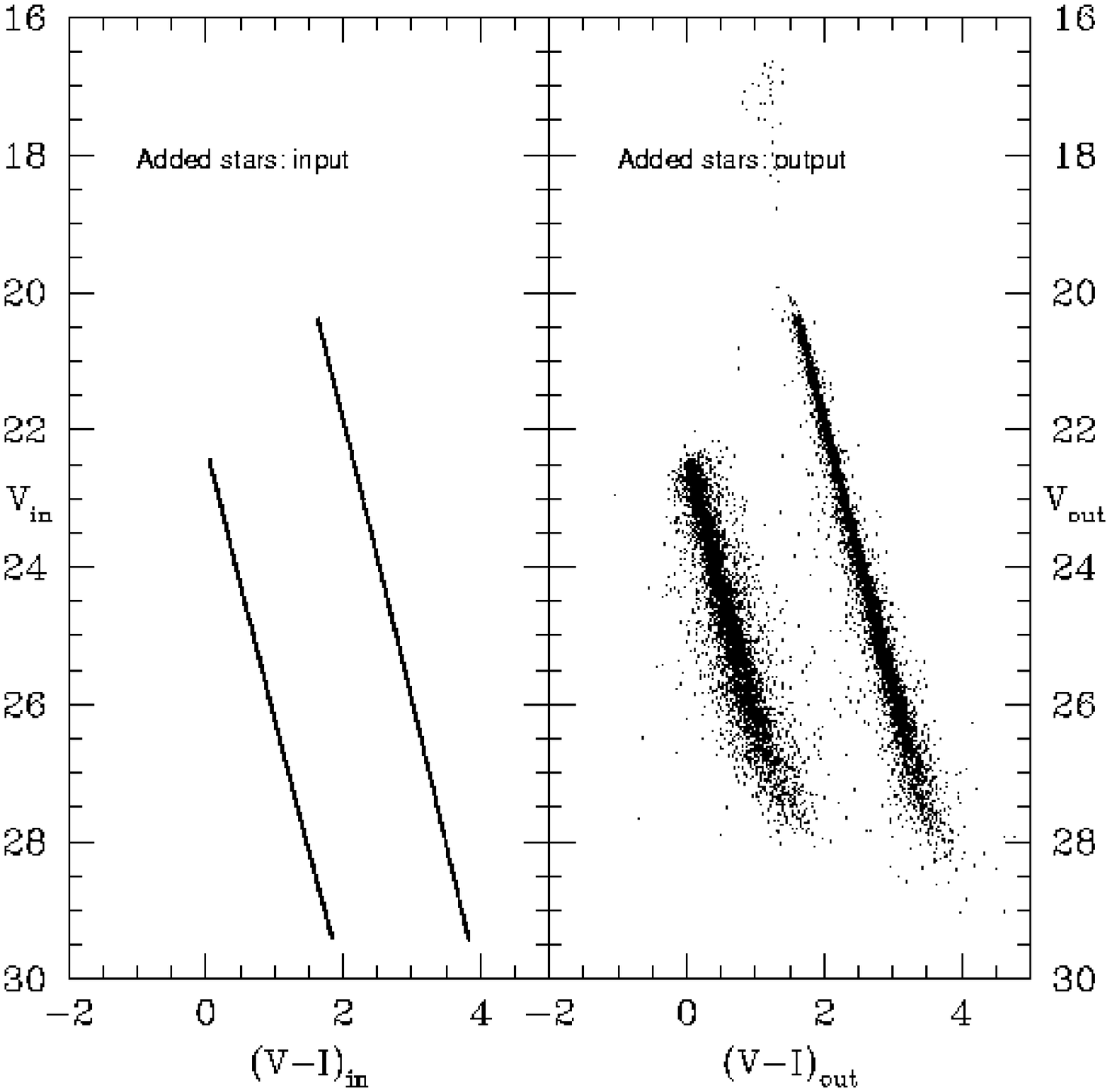,width=8.0cm,angle=0}
\caption[  ]{
The left-hand panel shows the artificial star  input sequences in V vs ${\rm
V-I}$.  A  total of $\sim 3  \times 10^4$ artificial  stars  were added. The
right-hand panel shows the sum of the recovered sequences from all fields.}
\end{figure}


\newpage


\begin{references}
 
\reference{Djor93}
Djorgovski, S., 1993 in Structure and Dynamics of Globular Clusters,
eds. S.G. Djorgovski \& G. Meylan (ASP:  San Francisco), 373. 

\reference{Fahl98}
Fahlman, G.G., Richer, H.B.,  Ibata, R.A., Stetson, P.B., Bell, R.A., Bolte, M.,
Bond, H.E., Harris, W.E., Hesser, J.E., Mandushev, G., Pryor, C. \& VandenBerg,
D.A., 1997, ApJ, submitted

\reference{Holtz95a}
Holtzman, J.A., \etal\  Hester, J., Casertano, S., Trauger, J.T.,
Watson, A.M, Ballester, G.E., Gilda, E., Burrows, C.J., 
Clarke, J.T., Crisp, D., Evans, R.W., Gallagher, J.S., Griffiths, R.E.,
Hoessel, J.G., Matthews, L.D., Mould, J.R., Scowen, P.A., Stapelfeldt, K.R.,
Westphal, J.A. 1995a, PASP 107, 156

\reference{Holtz95b}
Holtzman, J.A., Burrows, C.J., Casertano, S., Hester, J., Trauger, J.T.,
Watson, A.M. \& Worthey, G., 1995b, PASP 107, 1065.

\reference{Kelson96}
Kelson, D.D., Illingworth, G.D., Freedman, W.F., Graham, J.A., Hill, R.,
Madore, B.F., Saha, A., Stetson, P.B., Kennicut, R.C., Mould, J.R., 
Hughes, S.M., Ferrarese, L., Phelps, R., Turner, A., Cook, K.H., Ford, H.,
Hoessel, J.G. \& Huchra, 1996, ApJ 463, 26.

\reference{Richer95}
Richer, H.B., Fahlman, G.G., Ibata, R.A., Stetson, P.B., Bell, R.A., Bolte, M.,
Bond, H.E., Harris, W.E., Hesser, J.E., Mandushev, G., Pryor, C. \& VandenBerg,
D.A., 1995, ApJ 451, L17.

\reference{Richer97}
Richer, H.B., Fahlman, G.G., Ibata, R.A., Pryor, C., Bell, R.A., Bolte, M.,
Bond, H.E., Harris, W.E., Hesser, J.E., Holland, S., Ivanans, N., VandenBerg,
D.A., Wood, M., 1997, ApJ 484, 741

\reference{Stetson87}
Stetson, P.B., 1987, PASP 99, 613.
 
\reference{Stetson92}
Stetson, P.B., 1992, ``Further Progress in CCD Photometry,'' in {\it Stellar
Photometry --- Current Techniques and Future  Developments}, IAU Coll.\ 136,
eds.\ C.~J.~Butler and I.~Elliot, p.~291

\reference{Stetson94}
Stetson, P.B., 1994, PASP 106, 250.
 
\reference{Thompson90}
Thompson \etal 1990, unpublished photometry.


\end{references}
\end{document}